# Analyzing students' collaboratively solving spherical unit vector problems in upper-level E&M through a lens of shared resources

Ying Cao

*School of Education & Child Development, Drury University, 900 N. Benton Ave., Springfield, MO, 65802*

Brant E. Hinrichs

*Department of Physics and Chemistry, Drury University, 900 N. Benton Ave., Springfield, MO, 65802*

We are interested in better understanding ways that students collaborate to solve conceptual physics problems in the context of spherical unit vectors in upper-level E&M, especially problems that have been shown to be difficult for students to solve individually on their own, but which groups of students have been more successful at. Using think-aloud interviews with students in small groups, we ask them to solve together on a large whiteboard conceptual problems from this E&M context. The interviews were video and audio recorded, and qualitatively analyzed using an emergent coding method and the resources framework [4]. Through this analysis, we observed one common mechanism in all three group-interviews whereby students collaborated effectively: first one student activated a conceptual resource and expressed it, then another student took up that idea, and finally the whole group together used that idea to move forward with the problem. This mechanism exemplifies a newer framework: *shared resources* [12]. We further analyzed students' collaboration through the lens of shared resources and identified multiple instances. We propose that the shared resources construct could be a potential tool to help understand how students collaborate in solving conceptual physics problems. In this paper, we report our methodology and the results from one group interview to illustrate one shared resource we identified and the role it played in helping students collaboratively solve the conceptual problem in this context. Future work and implications for instruction are suggested.

## I. INTRODUCTION

The upper-level E&M course (i.e., based on Griffiths [1]) involves the extensive integration of vector calculus concepts and notation with abstract physics concepts like field and potential. We hope that students take what they have learned in their math classes and apply it to help represent and make sense of the physics. Previous work showed that physics majors at different levels (pre-E&M course, post-E&M course, first year graduate students) had great difficulty expressing position vectors using spherical unit vectors [2]. Since then we have developed a series of problems for students to work on and discuss in groups in class in order to help them make sense of these concepts [3].

Hinrichs has been conducting a long-term research project on students' understanding of spherical unit vectors in upper-level E&M. For the work in this paper, he recruited students to complete several concept tests individually and then interviewed them in small groups (2-3 students each). The motivation of conducting interviews was to possibly capture students' ideas and thinking processes that were not shown in the written tests. The reason for interviewing in groups was because, in his experience, due to the difficulty of the subject matter, individual students often got stuck, and also because students talking to each other often yield richer insights into what they are actually thinking in the moment.

Through our explorative analysis of the interview data, we noticed that in these group interviews, the students collaborated together to successfully find the answers to these conceptual problems that had been in general quite difficult for previous students to complete individually [2,3]. We were interested in learning more about the mechanisms of students' collaboration in the interviews and understanding in what ways their interactions had (or not) contributed to their success in solving these problems. Thus, our research question, which emerged through this data exploration, is: how do students collaborate on physics problems designed to make sense of non-Cartesian unit vectors in the context of E&M?

Since we were interested in examining students' detailed conceptual understandings of the topic, we adopted the resources framework [4] as our primary analytical lens[1]. This framework views student knowledge as existing in abundant, fine-grain-sized ideas, named *resources*, which can be activated due to a particular context. For example, "closer means stronger" is a resource that could be activated when someone tries to explain why they feel warmer when they sit nearer to a fireplace [7].

Previous work used the resources framework to identify resources and several groupings of resources of students' thinking about non-Cartesian unit vectors when describing position and velocity vectors in the context of upper-level mechanics through one-on-one interviews [8,9]. Our work adds to the literature [10, 11] by studying student problem solving using spherical unit vectors in the different physics context of upper-level E&M, and the different social context of small group think-aloud interviews.

Because we examine students' conceptual resources in the context of collaborative problem solving, we specifically apply a newer construct, *shared resources* [12], to look at both the activation and expression of resources, and the collective use of those resources by the group. This *shared* resources framework is built on the *resources* framework [4][2] but extended to a social context of learning [13]. We claim that one mechanism of students' collaborative problem solving in the context of our study can be explained through the lens of shared resources.

## II. THEORETICAL FRAMEWORK

A shared resource is a resource that is first activated by one person and expressed as an idea (in verbal, written, diagrammatic, etc. form), which is then taken up by another person in the same social context, and then collectively used by the group for the task at hand. In this case, the resource activated by the first person (a *private* resource) becomes a shared resource for the group.

For example, imagine two students were working together on a problem to find the direction of the magnetic field generated by a very long wire carrying a steady current that flows up. Suppose one of the students activated the resource "the right-hand rule" (a thumb-up gesture of the right hand, with the thumb aligned parallel to the wire and pointing in the direction of the current flow, and the circular four fingers indicating the direction of the B-field) that they learned from previous classes and expressed it to their partner as an idea either verbally, with a gesture, with a diagram, or in any combination of the three. The other student might not necessarily have had this resource activated, but on hearing or seeing the first student express it, felt the idea made sense and could help answer the problem. The second student then might go along with that idea, and the two students could both use the activated resource together to find the canonical direction of the magnetic field. In this case the resource "right-hand rule" became a shared resource.

In this paper, we adopted the shared resources framework to help us understand and articulate one mechanism students used to collaborate and solve a conceptual physics problem. In the following sections we describe how we collected and analyzed the data, report one instance of a shared resource in detail, and provide some possible implications for instruction.

---

[1] In this paper, we focus on conceptual resources—bits of students' understanding of physics topics, rather than other categories [5, 6].

[2] This idea of using the resources framework to explain social learning is not new, as it was briefly mentioned in a footnote in [4].

## III. METHODS

### A. Context

This study is part of a long-term project conducted by Hinrichs at his university. He alternately teaches a two-semester sequence of upper-level E&M with the one other faculty member in the physics program. The present paper focuses on analyzing video data from interviews conducted towards the end of the second semester with students then enrolled in the course. In that year, the other faculty member taught the course, while the interviews were conducted by Hinrichs. The instructional strategies applied in that E&M class were primarily traditional lectures, but the students had had at least one semester of experience with modeling, whiteboard work, coordinating multiple representations, and consensus building [14,15,16] in introductory physics with Hinrichs.

### B. Participants

Hinrichs announced to all the students then enrolled in the course that they were recruiting participants to be interviewed about course topics for this research project. Seven students volunteered to participate. Among them, six were physics majors, and one was a math major but also a physics minor. Six were identified as white male, and one was identified as an Asian female who was also an international student.

### B. The Task

A week before their interviews, the seven students were given three conceptual problems and asked to solve them on their own before their interview. Two of the three problems were previously published in [3]. The new one, which is the focus of this paper, is shown in Fig. 1. About a week later, the students were interviewed in three separate groups, which were pre-assigned by the interviewer. Students were grouped with those whom they seemed to feel comfortable with, based on the interviewer's perception of their small group interactions in previous classes. One group had a white male and the Asian female, one group had two white males, and one group had three white males. Different groups were interviewed on different days, but within a week of each other. Their individual written answers to the three conceptual problems were collected at the beginning of their interview. Then each group was given the same three problems in a particular order and asked to work together to solve them on a large white board at the front of the small classroom. The group with three students spent about an hour of concentrated time solving the problems for the interview. The other two groups spent about two hours each for their interview. We speculate that the group of three took less time because they had the only student who answered the entire Fig. 1 problem correctly on their own before the interview.

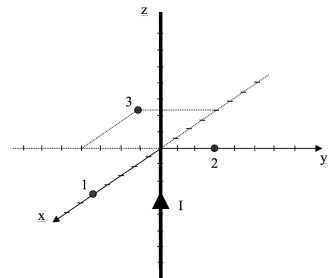

A very long wire lies along the z-axis (only a small portion of which is shown in the diagram below), and carries a steady current I that flows up.

(a) Please draw in the direction of the magnetic field vector, $\bar{B}$, at the following three points.
(b) Indicate what the direction of the magnetic field vector, $\bar{B}$, is at the three points, in terms of $\hat{r}, \hat{\theta}$ and $\hat{\varphi}$. Explain.

FIG. 1 The new concept problem given to students.

### C. Data Collection

Students participated in think-aloud interviews—i.e., they were asked to say everything out loud that they were thinking simultaneously as they were solving a problem [17, 18]. They were also instructed to work on a given problem until they felt they were done and then alert the interviewer. The interviewer generally let a group work by themselves, only occasionally gently prompting them with "What are you thinking?" if they became too quiet in expressing what they were thinking or doing. The interviews were both video and audio-recorded. Throughout the interview, photos were periodically taken of their written work on the whiteboard to help facilitate later analysis.

### D. Data Analysis

To begin to answer our research question, we conducted emergent coding—identifying themes that emerged from the ways students collaborated in their group problem solving—which eventually led us to adopt the construct of shared resources [12] to understand the moment-by-moment mechanism of their collaboration. Both authors participated in the coding process, which involved three rounds of analysis. Each round had its own focus, but taken all together they led to a progressively refined, thick description [19] of students' thinking and collaboration.

In the first round of analysis, we met together and watched the interview videos, just to get an overview of each group. We did not deliberately apply any framework while watching. Instead, we focused on jotting down the steps students had taken towards solving the problems. This round resulted in a set of descriptive notes outlining who did what at what time of the interview, whether or not the group had solved the particular problem, and whether there were crucial instances where students got stuck or had an a-ha moment. By the end of this round, we recognized that in each interview, there were multiple times when every group member contributed to the discussion of a part that they had gotten stuck on, and this collaborative discussion helped them figure it out and move forward.

We decided to look more closely at those instances to understand what exactly happened in their interaction to get them unstuck. Because we recognized that students were expressing rich conceptual understanding of the relevant topic at hand (in words, gestures, and visual representations on the white board), we decided to adopt the resources framework to examine their ideas more closely.

In the second round of analysis, we continued meeting and watching the videos together, with more frequent pausing and rewinding, to identify the conceptual resources activated and expressed by the students. We categorized these resources and reported our preliminary findings in a conference talk [20]. In this round we also realized that solely using the resources framework would not allow us to capture the role of student-student interactions and the way they collaboratively used resources. These two phenomena were extensive in all three interviews, and seemed to have played an important role in helping them solve the problem at hand. From there, we decided to adopt the shared resources framework [12] to conduct a third round of analysis.

In this final round, we first established a functional approach to identifying an instance of a shared resource. Each instance should include three parts in order: (1) "putting it down," where one person in the group activates a resource and expresses it as an idea, (2) "picking it up," where at least one other in the group takes up the idea by engaging with it instead of ignoring it; and (3) "using it," where everyone in the group uses the idea to make progress with the problem. We then used this approach to look in-depth at the first 30 min. of the interview with the group of Dave and Jules (these are pseudonyms; both are white males). We chose this group's interview for in-depth analysis because these two students were continuously verbalizing their thinking, giving us more direct evidence of their understanding, so that we would not have to infer as much. And we chose this section of the interview because they got very stuck solving part (b) of the Fig. 1 problem, but were able to figure it out all on their own. Only this group had both these features.

Applying our functional approach, we coded the first 30 min. of video sequentially in 10-min chunks. For each chunk, we separately identified instances of shared resources, then met together and checked for reliability, and then moved on to code the next chunk. After discussion together, we reached 75% agreement on the 12 shared resources identified in the first ten min., 93% agreement for 14 in the second ten min., and 100% agreement for 8 in the third ten min. We speculated that this increase of agreement over time could be, in part, because early in the interview, students activated various resources, but it was not clear whether and how they were going to use those ideas going forward. Therefore, the two authors had different interpre-tations. In the later chunks, such vagueness decreased, and the agreement increased.

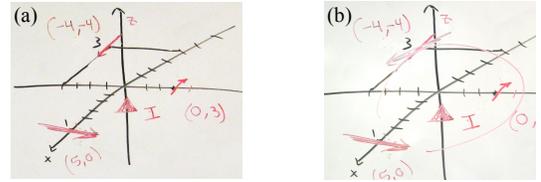

FIG. 2. Diagrams by Dave and Jules for part (a) of the Fig. 1 problem. (a) Red arrows show directions they think the B-field points at the three points. (b) The arrow at point 3 has been fixed.

## IV. RESULTS

We found 34 instances of shared resources throughout the first 30 min. of the interview, but because of limited space, we only describe one in detail here. Our description below first outlines the context right before the shared resource episode, then illustrates the three parts of it: "putting it down," "picking it up," and "using it," to find the canonical answer to part (b) of the Fig. 1 problem.

We chose this particular shared resource because the three part structure is typical, but their approach to solving the problem, and the productive representation they co-constructed by applying the shared resource was unique in Hinrichs' twelve years of experience.

Dave and Jules quickly solved part (a) of the Fig. 1 problem. Dave suggested the right hand rule, Jules readily agreed, applied it to a diagram of the problem that Dave had already drawn on the whiteboard, and drew red vectors to represent the B-field at the three points (Fig. 2a). Without mentioning it, the interviewer noticed their direction for point 3 was incorrect, and just asked them to make a top-down view of the problem as well. After working with that representation for a few minutes, Jules realized their error for point 3 and corrected it (Fig. 2b).

They then started solving part (b) of that problem. Previously, while Jules was just starting to draw Fig. 2a, Dave had skipped ahead and read part (b) out loud to Jules, who replied, in referring to $(\hat{r}, \hat{\theta}, \hat{\varphi})$, "*I don't even know what those are!*" Dave responded by recalling the definitions of $(r, \theta, \varphi)$ from calculus and drawing Fig. 3a to illustrate what he meant. Jules appeared satisfied: "*ok, yeah, that makes a lot more sense.*"

We argue that this is both the first part and second part of a shared resource. Dave does the first part: in response to Jules' quote of confusion above, Dave activates a resource, the definition of spherical coordinates $(r, \theta, \varphi)$ that he recalls from his previous calculus class, and expresses his idea verbally and diagrammatically. Jules does the second part with his second quote: he "picks up" the idea and readily agrees with it.

Next they quickly figured out and drew in $\hat{r}$ (in blue) on the diagram shown in Fig. 3b and moved on to try to make sense of $\hat{\theta}$. This is when the third part of the shared resource, "using it", starts. Dave drew a solid green arc from the +z-axis down to point 2, with an arrowhead at point 2 to indicate a clockwise direction, and labeled it with $\hat{\theta}$ (Fig. 3b). From now on we refer to this as a directed arc. The interviewer recognized that the drawing for $\hat{\theta}$ was incorrect, but did not intervene at this point. Jules asked why the beginning of the green directed arc started at that particular point on the +z axis and a discussion ensued where they referred back to and used the diagram of Fig. 3a (i.e., the original activated resource), and then drew another diagram (Fig. 4a), using that original resource to specifically work out $(r, \theta, \varphi)$ of point 2 at (0, 3, 0).

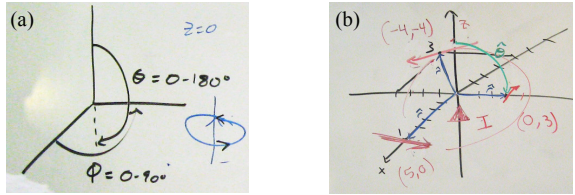

FIG. 3. Diagrams drawn by Dave. (a) The activated resource - his understanding of $(r, \theta, \varphi)$. (b) "Use" of the resource - the green directed arc labeled $\hat{\theta}$.

Jules finally agreed with the green directed arc, but then critiqued it, modified it, and extended its meaning to find the correct representation of $\hat{\theta}$:

Jules: *Ok, I see what you're saying…*
*Can a vector be curved like that?*
*The r-hat looks pretty good, but the theta-hat…*
(Dave draws Fig. 4a and explains his thinking again)
Jules: *Ok, then I agree with that but we need to redraw it because…I'm saying that this should be… dashed* (he dashes the green directed arc.)
*Until… we get right here* (points to y-axis)
*And it needs to point that way* (gesture unclear)
*Like, this* (points to green arc, Fig. 3b) *is NOT a unit vector, right here…*
Dave: *So you're saying it's not in the x-y plane, ok, yeah.*
Jules: *This is not a unit vector that you have in green.*
*This* (he draws a short purple arrow pointing down with its tail at point 2, Fig. 4b) *IS a unit vector!*
Dave: *Yeah.*

The transcript above of their process of dashing (by Jules) the directed green arc (drawn by Dave) and producing the purple arrow at point 2 (by Jules) provides evidence of both students collaboratively using the activated resource of Fig. 3a to figure out the accepted direction for $\hat{\theta}$.

Figure 4b also provides further evidence that the resource in Fig. 3a is a shared resource, as both students use it again in an effort to also figure out the direction of $\hat{\varphi}$ at point 2. Jules actively used the resource of Fig. 3a to draw the black dashed directed arc in the first quadrant of the x-y plane shown in Fig. 4b and together he and Dave figured out that $\hat{\varphi}$ must point in the direction of the red arrow that was already there. So then Jules drew in the short black ar-row pointing in the negative x-direction at point 2, which is the canonically correct answer. That Dave then went on to also use the same approach to correctly figure out $\hat{\theta}$ and $\hat{\varphi}$ for point 3 is additional evidence that both students were using the shared resource. Their success at this part of the problem is significant because previous work showed that finding the directions of $\hat{\theta}$ and $\hat{\varphi}$ was where most students struggled and were unable to answer correctly on their own [3].

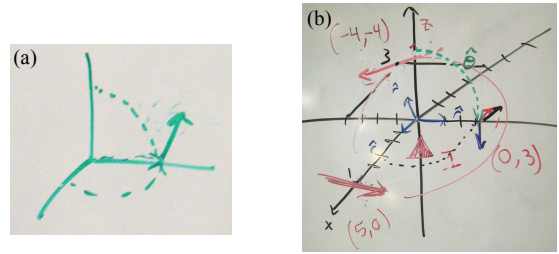

FIG. 4. (a) Use of resource again by Dave – diagram to explain how he thought about $(r, \theta, \varphi)$ for point 2. (b) Use of resource by Jules to find the correct representation for both $\hat{\theta}$ and $\hat{\varphi}$.

## DISCUSSION AND FUTURE WORK

Because visual representation was very important to thinking about and solving this problem, the shared resource we reported was a diagram, but, in principle, different kinds of problems might lend themselves to a verbal or textual shared resource instead. Shared resources are just one possible mechanism for modeling how students effectively collaborate together on problem solving. In the future we would like to explore how other models for this process, such as knowledge co-construction [21] and convergent conceptual change [22], compare and contrast with shared resources. Possible questions include: how common are shared resources in successful student collaboration?, what kinds of problems lend themselves to a shared resources framework analysis?, and more.

We end with two brief comments about possible implications for instruction. If instructors pay attention to how students share resources, they could facilitate both activation and sharing of known (from the literature) shared resources if students don't spontaneously activate them on their own. Lastly, the co-constructed representation of Fig. 4b might be a useful tool to explicitly teach students to help them better understand the meaning, use, and application of non-Cartesian unit vectors, especially $\hat{\theta}$ and $\hat{\varphi}$.

## ACKNOWLEDGMENTS

We thank Drury for professional development support, the students for their participation, and colleagues who provided feedback. Brant thanks Miki, Eads, and Fenn for their continued interest, encouragement, patience, and support, and the Kindness, Mercy and Grace of God.

[1] D. J. Griffiths, *Introduction to Electrodynamics*, 3rd Edition, Upper Saddle River, New Jersey: Prentice Hall, 1999.
[2] B. E. Hinrichs, Writing position vectors in 3-d space: a student difficulty with spherical unit vectors in intermediate E&M, *PERC 2010 Proceedings*, 173-6.
[3] B. E. Hinrichs, Student objections to and understanding of non-Cartesian unit vector notation in upper-level E&M, *PERC 2017 Proceedings*, pp 176-179.
[4] D. Hammer, A. Elby, R. E. Scherr, and E. F. Redish, Resources, framing, and transfer, *Transfer of learning from a modern multidisciplinary perspective* (2005): 89-120.
[5] D. Hammer, The variability of student reasoning, lecture 1: Case studies of children's inquiries, *The Proceedings of the Enrico Fermi Summer School in Physics*, Course CLVI (Italian Physical Society, 2004).
[6] R. E. Scherr and D. Hammer, Student behavior and epistemological framing: examples from collaborative active-learning activities in physics, C*ognition and Instruction,* **27**:2, 147-174 (2009).
[7] D. Hammer, Student resources for learning introductory physics, Am. J. Phys., **68** (S1), S52-S59 (2000).
[8] M. Vega, W. Christensen, B. Farlow, G. Passante, and M. Loverude, Student understanding of unit vectors and coordinate systems beyond Cartesian coordinates in upper division physics courses, *PERC 2016 Proceedings*, 364-367.
[9] B. Farlow, M. Vega, M. E. Loverude, and W. M. Christensen, Mapping activation of resources among upper division physics students in non-Cartesian coordinate systems: A case study, Phys. Rev. Phys. Educ. Res., **15**, 020125 (2019).
[10] C. Singh, Impact of peer interaction on conceptual test performance, Am. J. Phys., **73**(5), 446-451 (2005).
[11] A. Pawlak, P. W. Irving, and M. D. Caballero, Development of the modes of collaboration framework, Phys. Rev. Phys. Educ. Res. **14,** 010101, (2018).
[12] Y. Cao, and M. D. Koretsky, Shared resources: Engineering students' emerging group understanding of thermodynamic work, J. Eng. Educ.*,* **107**:4, 656-689, (2018).
[13] L. S. Vygotsky, *Mind in society: The development of higher psychological processes,* Harvard university press, 1980.
[14] B. E. Hinrichs, Using the system schema representational tool to promote student understanding of Newton's third law, *PERC 2004 Proceedings*, 117–120.
[15] B. E. Hinrichs, Sharp initial disagreements then consensus in a student led whole-class discussion, *PERC 2013 Proceedings*, 181-184.
[16] D. T. Brookes, B. E. Hinrichs, and J. L. Nass, Social positioning correlates with consensus building in two contentious large-group meetings, *2019 PERC Proceedings*.
[17] K. Ericsson, K., and H. Simon, *Protocol Analysis: Verbal Reports as Data* (2nd ed.), Boston: MIT Press, 1993.
[18] J.P. Leighton, Two types of think aloud interviews for educational measurement: Protocol and verbal analysis, paper presented at the annual meeting of the National Council on Measurement in Education, San Diego, CA., 2009.
[19] C. Geertz, *Thick Description: Toward an Interpretive Theory of Culture*, 1973.
[20] Y. Cao and B. Hinrichs, Students' conceptual resources for spherical unit vectors in upper-division E&M, contributed talk at *AAPT 2019 summer meeting.*
[21] S. Jacoby, & E. Ochs, Co-construction: An introduction, Res. Lang. Soc. Interact., **28**(3),171–184, (1995).
[22] J. Roschelle, Learning by collaborating: convergent conceptual change, JRST **2**(3), 235-276, (1992).